%% file: Formatting-Instructions-LaTeX-2026.tex
\documentclass[letterpaper]{article} 
\usepackage{aaai2026}  
\usepackage{times}  
\usepackage{helvet}  
\usepackage{courier}  
\usepackage[hyphens]{url}  
\usepackage{graphicx} 
\usepackage{natbib}  
\usepackage{caption} 
\usepackage{booktabs} 
\usepackage{multirow} 
\usepackage{amsmath} 
\usepackage{algorithm}
\usepackage{algorithmic}
\usepackage{xcolor}

\usepackage{newfloat}
\usepackage{listings}
\DeclareCaptionStyle{ruled}{labelfont=normalfont,labelsep=colon,strut=off} 
\floatstyle{ruled}
\newfloat{listing}{tb}{lst}{}
\floatname{listing}{Listing}
\title{PersonaAct: Simulating Short-Video Users with Personalized Agents\\ for Counterfactual Filter Bubble Auditing}
\author{
    Shilong Zhao\textsuperscript{\rm 1,\rm 2},
    Qinggang Yang\textsuperscript{\rm 1,\rm 2},
    Zhiyi Yin\textsuperscript{\rm 1}\thanks{Corresponding Author.},
    Xiaoshi Wang\textsuperscript{\rm 1},\\
    Zhenxing Chen\textsuperscript{\rm 1,\rm 2},
    Du Su\textsuperscript{\rm 1},
    Xueqi Cheng\textsuperscript{\rm 1}
}
\affiliations{
    \textsuperscript{\rm 1} State Key Lab of AI Safety, Institute of Computing Technology, CAS, Beijing, China \\
    \textsuperscript{\rm 2} University of Chinese Academy of Science, Beijing, China \\
    zhaoshilong0108@126.com, \{yangqinggang25e, yinzhiyi, wangxiaoshi\}@ict.ac.cn, \\chenzhenxing2021@gmail.com, \{sudu, cxq\}@ict.ac.cn
}

\usepackage{bibentry}

\begin{document}

\maketitle

\begin{abstract}
\IfFileExists{common/0abs_cr.tex}{
  \input{common/0abs_cr}
}{
  \input{common/0abs}
}
\end{abstract}

\begin{links}
    \link{Code \& Datasets}{https://github.com/Silung/PersonaAct}
    \link{Demo}{https://huggingface.co/spaces/Silung/PersonaActInterviewAgent}
\end{links}

\IfFileExists{common/1intro_cr.tex}{
\input{common/1intro_cr}
}{
  \input{common/1intro}
}

\IfFileExists{common/2related_cr.tex}{
  \input{common/2related_cr}
}{
  \input{common/2related}
}

\IfFileExists{common/3dataset_cr.tex}{
  \input{common/3dataset_cr}
}{
  \input{common/3dataset}
}

\IfFileExists{common/4method_cr.tex}{
  \input{common/4method_cr}
}{
  \input{common/4method}
}

\IfFileExists{common/5exp_cr.tex}{
  \input{common/5exp_cr}
}{
  \input{common/5exp}

}

\IfFileExists{common/6future_cr.tex}{
  \input{common/6future_cr}
}{
  \input{common/6future}
}

\IfFileExists{common/7con_cr.tex}{
  \input{common/7con_cr}
}{
  \input{common/7con}

}

\bibliography{common/aaai2026}

\end{document}

%% file: common/0abs.tex
Short-video platforms rely on personalized recommendation, 
raising concerns about filter bubbles that narrow content exposure. 
Auditing such phenomena at scale is challenging 
because real user studies are costly and privacy-sensitive, 
and existing simulators fail to reproduce realistic behaviors due to their reliance on textual signals and weak personalization.
We propose \textbf{PersonaAct}, a framework for simulating short-video users 
with persona-conditioned multimodal agents trained on real behavioral traces for auditing filter bubbles in breadth and depth. 
PersonaAct synthesizes interpretable personas through automated interviews 
combining behavioral analysis with structured questioning, 
then trains agents on multimodal observations using supervised fine-tuning 
and reinforcement learning. 
We deploy trained agents for filter bubble auditing and evaluate bubble breadth via content diversity and bubble depth via escape potential. 
The evaluation demonstrates substantial improvements in fidelity over generic LLM baselines, 
enabling realistic behavior reproduction. 
Results reveal significant content narrowing over interaction. 
However, we find that Bilibili demonstrates the strongest escape potential.
We release the first open multimodal short-video dataset and code 
to support reproducible auditing of recommender systems.

%% file: common/1intro_cr.tex
\section{Introduction}

Short-video platforms have become dominant content distribution channels, 
with personalized recommendation algorithms shaping billions of users' information diets. 
While these systems improve engagement, 
they risk creating filter bubbles \cite{pariser2011filter,flaxman2016filter} 
that progressively narrow content exposure, 
potentially limiting informational diversity and reinforcing existing preferences. 
Auditing such algorithmic effects is crucial for understanding 
their societal impact and ensuring responsible deployment.

\begin{figure}[t]
    \centering
    \includegraphics[width=0.95\linewidth]{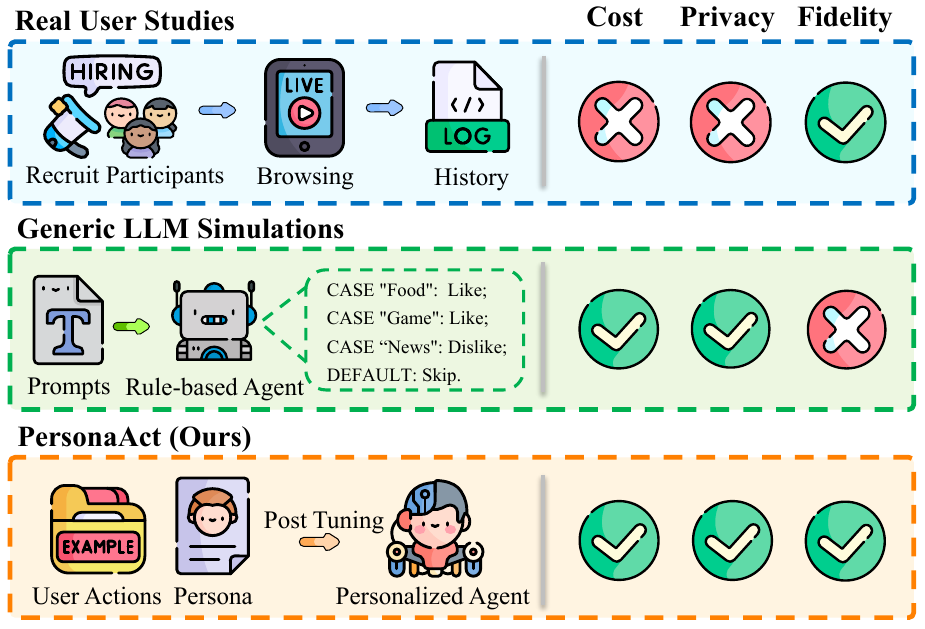}
    \caption{RecSys auditing approaches: cost, privacy, fidelity trade-offs.}
    \label{fig:motivation}
\end{figure}

However, auditing filter bubbles at scale faces fundamental challenges. 
Real-user studies provide authentic behavioral data but incur high costs 
and raise privacy concerns that limit scalability. 
Alternative approaches using sockpuppet accounts \cite{Boeker_2022,Haroon_2023} 
lack the naturalness required for realistic auditing. 
Existing simulators offer a scalable alternative but suffer from critical limitations. 
Rule-based approaches lack adaptability to diverse user behaviors, 
while recent LLM-based agents \cite{wang2024userbehaviorsimulationlarge,zhang2024llmpoweredusersimulatorrecommender} 
rely predominantly on textual signals, 
failing to capture the multimodal nature of short-video engagement 
where visual and audio cues drive viewing decisions. 
Moreover, these simulators employ weak personalization through generic prompting, 
unable to reproduce individual behavioral nuances essential for reliable auditing. 
As Figure~\ref{fig:motivation} illustrates, 
existing approaches face inherent trade-offs between cost, privacy, and behavioral fidelity.

We propose \textbf{PersonaAct}, a framework for simulating short-video users 
with persona-conditioned multimodal agents that achieve high behavioral fidelity 
while maintaining scalability and privacy. 
PersonaAct addresses the personalization gap through automated persona synthesis 
that combines behavioral feature extraction with structured interviews, 
producing interpretable user profiles grounded in real browsing patterns. 
These personas then condition multimodal agents trained via supervised fine-tuning 
and reinforcement learning to predict user actions from video frames and audio content. 
The resulting agents enable controlled counterfactual experiments 
for auditing filter bubble dynamics across multiple platforms.

Evaluation on two distinct personas demonstrates substantial improvements 
in simulation fidelity over generic LLM baselines. 
Deploying these agents for filter-bubble auditing on Bilibili, Douyin, and Kuaishou 
reveals significant content narrowing across platforms, 
with diversity declining over sustained interactions. 
Counterfactual analysis further uncovers platform-dependent algorithmic inertia, 
with varying levels of resistance to behavioral changes across services.

This work makes three primary contributions:
\begin{itemize}
    \item We release the first open multimodal dataset for short-video simulation 
    with interview-derived personas capturing diverse behavioral patterns.
    \item We develop persona-conditioned multimodal agents that substantially improve 
    behavioral simulation fidelity through automated persona synthesis and reinforcement learning.
    \item We conduct counterfactual auditing across major platforms, 
    quantifying filter bubbles through content diversity and escape potential metrics.
\end{itemize}

%% file: common/1intro.tex
\section{Introduction}

Short-video platforms rely on personalized recommender systems 
that shape billions of users' information diets. 
While improving engagement, 
these systems risk creating filter bubbles \cite{pariser2011filter,flaxman2016filter} 
that narrow content exposure and limit informational diversity. 
Auditing such effects is crucial for understanding their societal impact.

\begin{figure}[t]
    \centering    \includegraphics[width=0.95\linewidth]{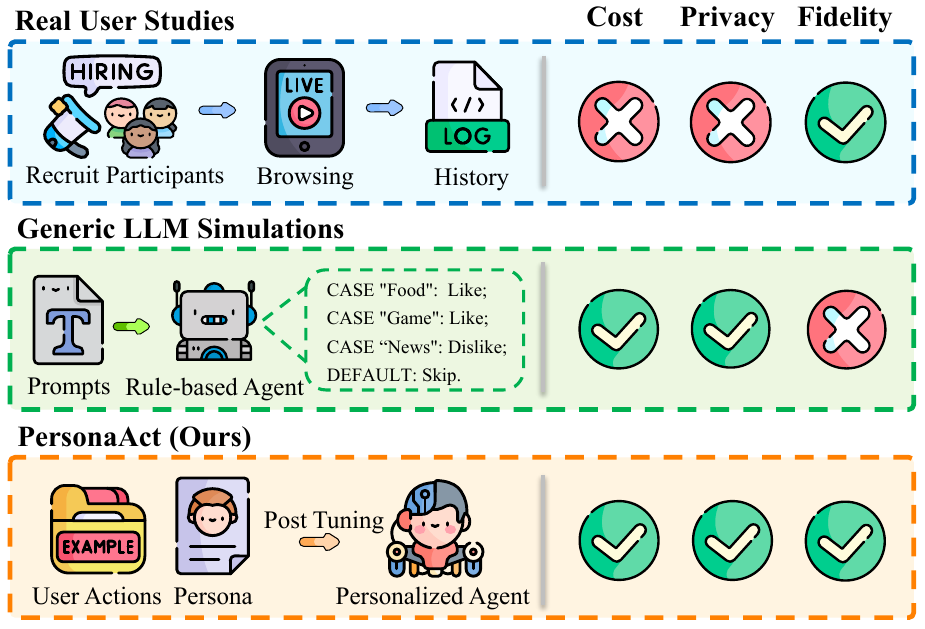}
    \caption{RecSys auditing approaches: cost, privacy, fidelity trade-offs.}
    \label{fig:motivation}
\end{figure}

However, auditing filter bubbles at scale faces fundamental challenges. 
Real user studies provide authentic data but incur high costs and privacy concerns. 
Alternative approaches using sockpuppet accounts \cite{Boeker_2022,Haroon_2023} 
or simulators offer scalability but suffer from critical limitations. 
Rule-based simulators lack adaptability, 
while recent LLM-based agents \cite{wang2024userbehaviorsimulationlarge,zhang2024llmpoweredusersimulatorrecommender} 
rely on textual signals and weak personalization, 
failing to capture the multimodal nature of short-video engagement 
where visual and audio cues drive decisions. 
As Figure~\ref{fig:motivation} shows, 
existing approaches face trade-offs between cost, privacy, and fidelity.

We propose \textbf{PersonaAct}, a framework for simulating short-video users 
with persona-conditioned multimodal agents. 
PersonaAct synthesizes interpretable personas through automated interviews 
combining behavioral analysis with structured questioning, 
then trains agents via supervised fine-tuning and reinforcement learning 
to predict actions from video frames and audio content. 
The resulting agents enable controlled counterfactual experiments 
for auditing filter bubble dynamics across platforms.

Evaluation demonstrates substantial fidelity improvements over generic LLM baselines. 
Deploying agents on Bilibili, Douyin, and Kuaishou 
reveals significant content narrowing and platform-dependent algorithmic inertia.

\noindent\textbf{Contributions.}
\begin{itemize}
    \item We develop persona-conditioned multimodal agents that substantially improve 
    simulation fidelity through automated persona synthesis and reinforcement learning.
    \item We provide a comprehensive auditing framework that balances breadth and depth dimensions,
    achieving cost-efficient, privacy-preserving, and authentic filter bubble analysis at scale.
    \item We release the first open multimodal dataset for short-video simulation 
    with interview-derived personas.
\end{itemize}

Code and dataset are available on GitHub\footnote{\hbox{\url{https://anonymous.4open.science/r/PersonaAct-9B3E}}}, 
and the Interview Agent demo can be experienced on HuggingFace\footnote{\hbox{\url{https://huggingface.co/spaces/Silung/PersonaActInterviewAgent}} (Empty Account)}.

%% file: common/2related.tex
\section{Related Work}

\subsection{LLM Agents for User Simulation in RS}
Traditional recommender simulations relied on RL or cognitive models \cite{ie2019recsimconfigurablesimulationplatform,shi2018virtualtaobaovirtualizingrealworldonline}. Recently, LLM-based generative agents \cite{park2023generativeagentsinteractivesimulacra} 
have been adapted to recommender systems for user simulation. 
RecAgent \cite{wang2024userbehaviorsimulationlarge} and 
Agent4Rec \cite{zhang2024llmpoweredusersimulatorrecommender} 
simulate user-item interactions from textual logs, 
while SimUser \cite{bougie2025simusersimulatinguserbehavior} aligns personas with historical behavior 
and Customer-R1 \cite{wang2025customerr1personalizedsimulationhuman} 
uses GRPO to condition agents on survey-derived profiles.

However, these text-centric approaches \cite{wu2024surveylargelanguagemodels} overlook multimodal cues essential for short-video engagement and rely heavily on static one-shot prompting. To bridge these gaps, PersonaAct synthesizes dynamic personas via multimodal alignment and iterative refinement.

\subsection{Filter Bubble}

Auditing filter bubbles \cite{flaxman2016filter,ross2022echo} 
traditionally relies on sockpuppet accounts \cite{Boeker_2022, Mosnar_2025, Haroon_2023}, 
which lack scalability and behavioral realism. 
Recent work such as ClipMind \cite{gong2025clipmind} and 
SimTok \cite{sukiennik2025simulatingfilterbubbleshortvideo} 
employ multimodal AI and LLM agents for auditing, yet remain constrained by coarse textual profiles.
Prior work~\cite{feng2025quantifying} proposes Bubble Escape Potential (BEP) 
by comparing independent users with contrasting behaviors.  
We extend this concept by proposing a sequential counterfactual framework to measure algorithmic inertia, quantifying the dynamic difficulty of escaping established bubbles beyond static comparisons.

%% file: common/3dataset.tex
\section{Dataset: Short-Video User Actions}
\label{sec:dataset}

\begin{figure}[h]
\centering
\includegraphics[width=0.7\linewidth]{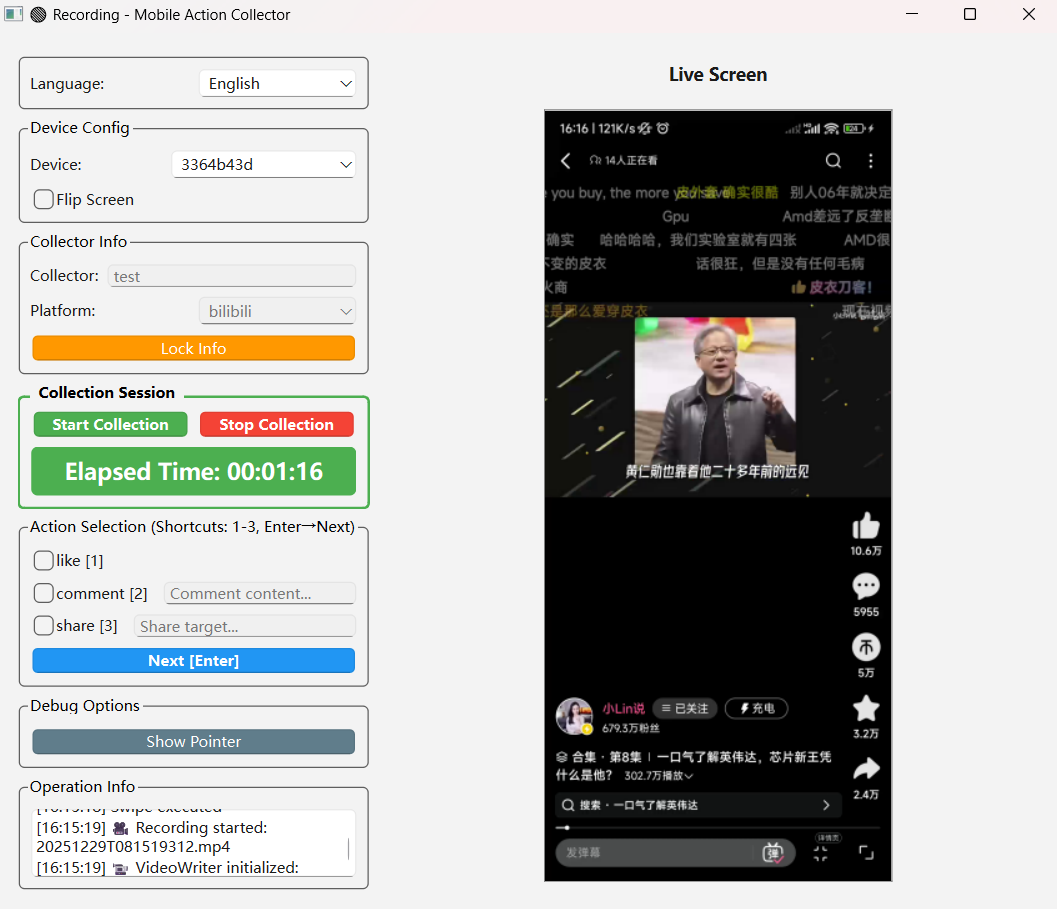}
\caption{Data collection interface mirroring mobile device screen.}
\label{fig:annotator}
\end{figure}

We collect a multimodal dataset capturing authentic short-video browsing behavior. 
As figure~\ref{fig:annotator} shows, our ADB-based recording tool synchronizes video frames, audio transcripts, metadata, 
and user actions including \texttt{watch}, \texttt{like}, \texttt{comment}, and \texttt{share} 
with precise timestamps.
Participants browsed on their regularly-used platforms to ensure authentic behaviors.
We collected data with informed consent and anonymized all recordings.

\subsection{Dataset Statistics}

Our dataset contains 1,719 video-interaction samples from 2 users across 27 browsing sessions. 
We adopt session-based temporal splitting to prevent leakage, 
with the first 80\% for training, middle 10\% for validation, and last 10\% for test, 
ensuring temporal ordering within sessions.

\begin{table}[h]
\centering
\caption{Dataset statistics}
\resizebox{\linewidth}{!}{%
\begin{tabular}{lccccc}
\toprule
\textbf{Persona} & \textbf{Samples} & \textbf{Sessions} & \textbf{Watch} & \textbf{Like} & \textbf{Comment/Share} \\
\midrule
Persona A & 939 & 8 & 936 (99.7\%) & 58 (6.2\%) & 2 (0.2\%) \\
Persona B & 780 & 19 & 780 (100\%) & 54 (6.9\%) & 4 (0.5\%) \\
\midrule
\textbf{Total} & \textbf{1,719} & \textbf{27} & \textbf{1,716 (99.8\%)} & \textbf{112 (6.5\%)} & \textbf{6 (0.3\%)} \\
\midrule
\multicolumn{6}{l}{\textbf{Watch Duration (seconds)}} \\
Persona A & \multicolumn{5}{l}{Mean ± Std: 7.8 ± 10.3; Median: 5.0; Range: [0.5, 120.6]} \\
Persona B & \multicolumn{5}{l}{Mean ± Std: 21.9 ± 40.7; Median: 7.3; Range: [0.8, 436.8]} \\
\bottomrule
\end{tabular}%
}
\label{tab:dataset}
\end{table}

\textbf{Key Characteristics.} 
The dataset exhibits extreme action sparsity typical of short-video browsing. 
While all samples include watch durations, likes occur in only 6.5\% and comments/shares in 0.3\%.
The two users show distinct personas with different engagement patterns, 
spanning entertainment, lifestyle, knowledge, sports, and music categories.
This motivates our focus on watch duration prediction as the primary evaluation metric.
We release the dataset, collection tool, and preprocessing scripts.

%% file: common/4method.tex
\section{Method: PersonaAct}

\begin{figure*}[t]
\centering
\includegraphics[width=0.98\textwidth]{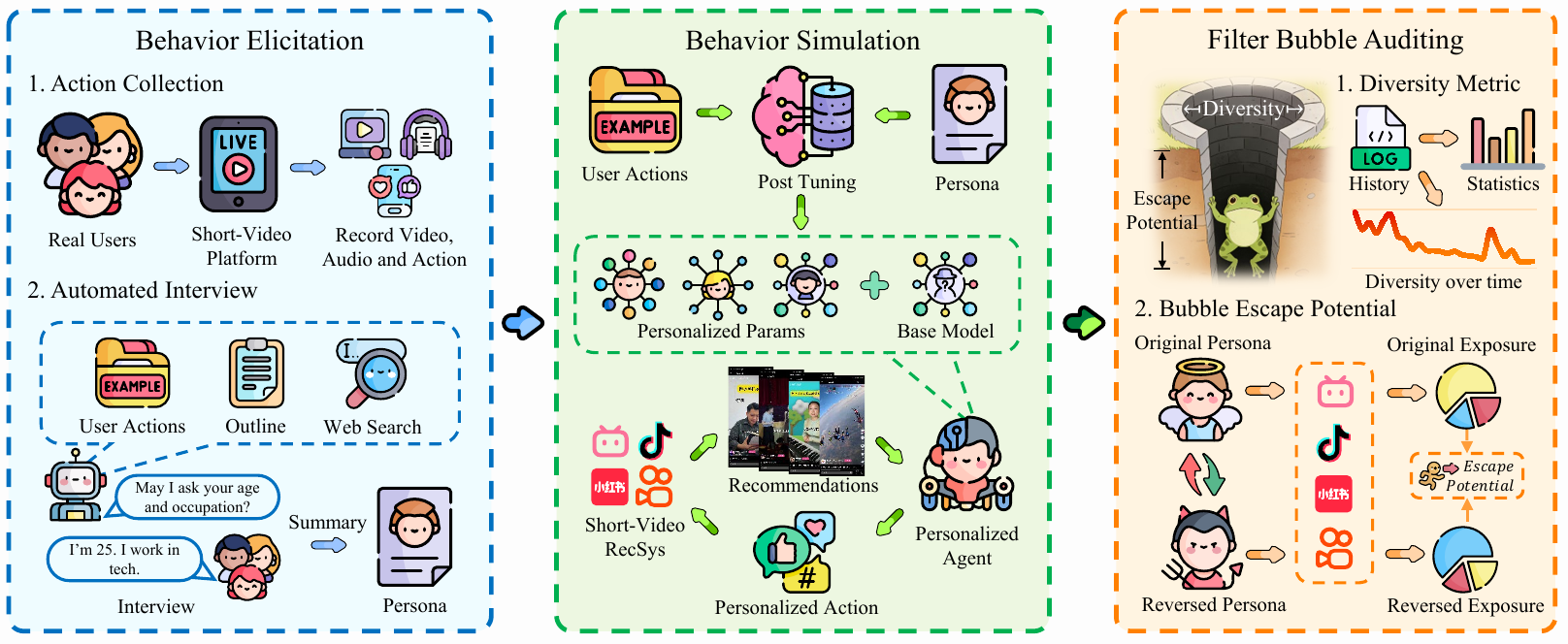}
\caption{Overview of PersonaAct. We first collect multimodal browsing traces from real users. An automated interview then integrates behavioral analysis with structured questioning to produce an interpretable persona. Next, we train a persona-conditioned agent to simulate user actions from multimodal context. Finally, the trained agents are deployed for filter bubble auditing by tracking content diversity and escape potential over time.}
\label{fig:framework}
\end{figure*}

PersonaAct generates personalized user agents by combining 
automated persona elicitation and persona-conditioned behavior learning, 
which are then used for filter bubble auditing. 
Figure~\ref{fig:framework} summarizes the overall workflow.

\subsection{Automated Persona Interview}

\begin{figure}[t]
\centering
\includegraphics[width=\linewidth]{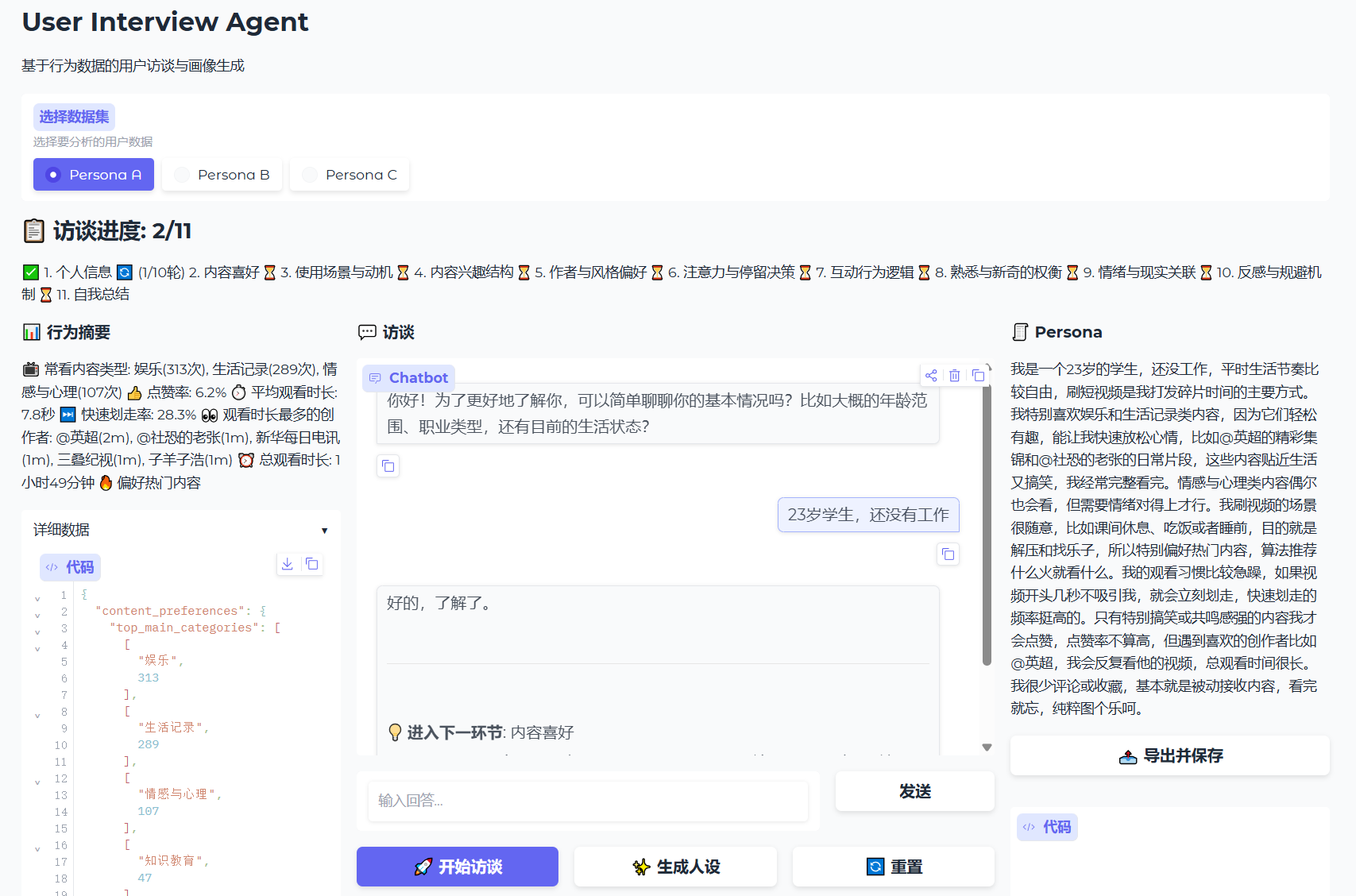}
\caption{Automated Persona Interview example.}
\label{fig:interview-agent}
\end{figure}

Behavior logs record \textit{what} users did, but provide limited evidence about \textit{why}. 
We design an Automated Persona Interview Agent that conducts structured conversations with users 
to elicit preferences, motivations, and decision-making tendencies, as shown in figure~\ref{fig:interview-agent}.

Given a logged trajectory $\tau=\{(o_t,a_t)\}_{t=1}^{T}$, 
the agent produces a persona profile $p=g_{\psi}(z,\mathcal{D})$ 
by combining behavioral features $z$ and dialogue responses $\mathcal{D}$.

\textbf{Data-Driven Questioning.} Before the interview, 
the behavior analyzer extracts features from session logs including 
category distributions, liked categories, like/comment/share rates, 
watch durations, creator preferences, temporal habits, and engagement metrics. 
It also identifies long-watched, quick-skipped, and liked videos as representative examples. 
During the interview, questions are dynamically generated as $q_j \sim \pi_I(\cdot\mid z,\mathcal{D}_{<j})$, 
adapting to behavioral patterns and prior responses to elicit 
explanations beyond surface-level preferences.

\textbf{Structured Outline.} The agent follows a configurable interview outline with 
predefined dimensions like  usage context, content preferences, creator affinity, 
attention mechanisms, engagement logic, exploration tendencies. 
Each dimension specifies goals and question directions. 
The agent maintains section-level context and transitions when section 
objectives are met, ensuring comprehensive yet focused coverage.

\textbf{Persona Synthesis.} 
The agent synthesizes interview responses into a persona profile 
that integrates self-narrative summaries of viewing motivations and preferences 
with personality traits (emotion regulation, novelty tolerance) 
and behavioral statistics (like rate, watch duration patterns), 
serving as conditioning input for behavior simulation.

\subsection{Behavior Simulation}
\label{sec:behavior-sim}

We formulate behavior simulation as learning a persona-conditioned 
policy $\pi_\theta(a_t\mid p,o_t)$ that predicts actions from 
multimodal observations and persona profiles. 
The action space comprises $\{\texttt{watch}, \texttt{like}, \texttt{comment}, \texttt{share}\}$ with associated watch durations.
We employ a two-stage training regime: supervised fine-tuning (SFT) 
establishes behavioral manifolds from demonstrations, followed by 
Group Relative Policy Optimization (GRPO) that refines the policy 
via a reward function $R=R_{\text{action}}+R_{\text{duration}}+R_{\text{format}}$.
$R_{\text{action}}$ uses F1-based matching for discrete actions 
(\texttt{like}, \texttt{comment}, \texttt{share}), 
$R_{\text{duration}}$ applies duration-sensitive scoring for \texttt{watch} actions:
\begin{equation}
\begin{aligned}
R_{\text{watch}}=1-\min\left(1,(|d-\hat{d}|)/d\right),
\end{aligned}
\end{equation}
where $d$ and $\hat{d}$ denote ground-truth and predicted durations,
and $R_{\text{format}}$ ensures parseable outputs.

\subsection{Counterfactual Filter Bubble Auditing}
\label{sec:filter-bubble-auditing}

We propose a counterfactual auditing framework to quantify filter bubble severity through controlled persona-driven interactions. 
Our approach measures two complementary dimensions: exposure diversity (bubble breadth) and escape potential under counterfactual engagement (bubble depth).

\textbf{Bubble Breadth via Fresh Account Deployment.}
To assess content diversity, we deploy trained agents on fresh accounts and 
track the number of distinct categories within sliding exposure windows 
(category entropy exhibits similar trends).

\textbf{Bubble Depth via Reversed-Persona Counterfactual.}
To quantify escape potential, we employ a contrastive counterfactual protocol. 
Unlike bubble breadth measurement, this approach operates on \emph{already-adapted accounts}, reducing deployment costs.
For persona $p$, we conduct two sequential interaction phases of equal duration.
In the \emph{cultivation phase}, the agent interacts under $p$, allowing the recommender to adapt.
In the \emph{counterfactual phase}, we deploy a reversed persona $p'$ that inverts engagement signals via quantile reversal. 
For predicted duration $\hat{d}_t$ at quantile $q$ in $p$'s training distribution, we sample $d_t' \sim F^{-1}(1-q)$, ensuring $P(d<\hat{d}_t) = P'(d>d_t')$.
We then measure the Bubble Escape Potential (BEP) based on the Jensen-Shannon divergence between content category distributions in the two phases:
\begin{equation}
\begin{aligned}
\mathrm{BEP}(p) &= \mathrm{JS}(P_p \parallel P_{p'}) \\
&= \frac{1}{2}\mathrm{KL}(P_p \parallel M) + \frac{1}{2}\mathrm{KL}(P_{p'} \parallel M),
\end{aligned}
\end{equation}
where $M=\frac{1}{2}(P_p+P_{p'})$ is the mixture distribution, and $P_p$ and $P_{p'}$ are the empirical category distributions for persona $p$ and $p'$, respectively. We use base-2 logarithm in computing JS divergence, yielding BEP$\in[0,1]$. Higher BEP indicates greater distributional divergence, meaning the platform responds more readily to behavioral changes—revealing stronger escape potential and weaker algorithmic inertia.

%% file: common/5exp.tex
\section{Experiments}

We organize our experiments around two research questions:

\noindent\textbf{RQ1:} Can persona-driven LLM agents realistically simulate user behaviors in short-video browsing?

\noindent\textbf{RQ2:} How severe are filter bubbles in short-video platforms, and can users escape them by changing behaviors?

\subsection{Implementation Details}

We fine-tune Qwen2.5-VL-7B-Instruct~\cite{qwen2.5-VL} using LoRA~\cite{hu2022lora} (rank=64, $\alpha$=128 for SFT; rank=8, $\alpha$=32 for GRPO) with learning rate 1e-5, batch size 1, gradient accumulation 2. 
GRPO uses 8 generations per sample with temperature 1.0 and KL coefficient $\beta$=0.01.
All experiments were conducted using NVIDIA H100 GPUs and controlled the same training steps for all methods.

\subsection{Simulation Fidelity (RQ1)}

We evaluate the ability of persona-driven agents to reproduce realistic user behaviors 
by comparing SFT, GRPO, and SFT+GRPO against LLM Sim baselines. 
Due to extreme action sparsity, we focus on watch duration prediction, reporting 
SMAPE (Symmetric Mean Absolute Percentage Error) and MAE (Mean Absolute Error). 
The SMAPE is defined as:
\begin{equation}
\begin{aligned}
\text{SMAPE} = \frac{1}{n}\sum\frac{|y_i-\hat{y}_i|}{(|y_i|+|\hat{y}_i|)/2}
\end{aligned}
\end{equation}
The SMAPE normalizes errors by video length, while the MAE captures absolute error magnitude. 
This dual reporting ensures balanced evaluation across 
heterogeneous video-length distributions.

\begin{table}[h]
\centering
\caption{Simulation fidelity}
\label{tab:fidelity}
\begin{tabular}{lcccc}
\toprule
\textbf{Persona} & \textbf{Method} & \textbf{SMAPE} & \textbf{MAE (s)} \\
\midrule
\multirow{4}{*}{Persona A} & LLM Sim & 1.161 & 6.69 \\
 & SFT & 0.683 & 5.61 \\
 & GRPO & 1.305 & 6.26 \\
 & SFT+GRPO & \textbf{0.617} & \textbf{5.10} \\
\midrule
\multirow{4}{*}{Persona B} & LLM Sim & 1.357 & 19.09 \\
 & SFT & 0.969 & 23.81 \\
 & GRPO & 1.218 & 20.40 \\
 & SFT+GRPO & \textbf{0.860} & \textbf{19.08} \\
\bottomrule
\end{tabular}%
\end{table}

\textbf{Observations.} SFT+GRPO consistently achieves the lowest error 
across both personas. SFT excels at imitation but lacks adaptability, 
while GRPO enables exploration but risks divergence. 
The two-stage approach anchors exploration within 
SFT-initialized manifolds for optimal calibration.

\textbf{Ablation Study.} 
We ablate persona conditioning and behavioral tuning to isolate 
their contributions. As Table~\ref{tab:ablation_persona} shows, 
removing persona conditioning increases the error 
from 0.617 to 0.633 (2.6\%), while removing behavioral tuning 
increases the error from 0.617 to 1.161 (88\%), 
indicating the essential role of both persona conditioning and 
behavioral tuning.

\begin{table}[h]
\centering
\caption{Ablation study on Persona A}
\label{tab:ablation_persona}
\begin{tabular}{lcc}
\toprule
\textbf{Method} & \textbf{SMAPE} & \textbf{MAE (s)} \\
\midrule
PersonaAct & \textbf{0.617} & \textbf{5.10} \\
-Persona & 0.633 & 5.15 \\
-Tuning & 1.161 & 6.69 \\
-Both & 1.236 & 6.96 \\
\bottomrule
\end{tabular}%
\end{table}

\subsection{Filter Bubble Auditing (RQ2)}

We deploy persona-driven agents to audit filter bubble formation 
in online short-video platforms including Bilibili, Douyin (TikTok) and 
Kuaishou, which collectively serve over 1.8 billion monthly active users.
Filter bubbles manifest through 
two complementary dimensions: 
\emph{Bubble Breadth} refers to the diversity of content categories 
a user is exposed to, while \emph{Bubble Depth} reflects 
how easily users can escape established content corridors. 
We quantify these with content diversity and 
Bubble Escape Potential (BEP), respectively.

\begin{figure}[htbp]
\centering
\includegraphics[width=0.85\linewidth]{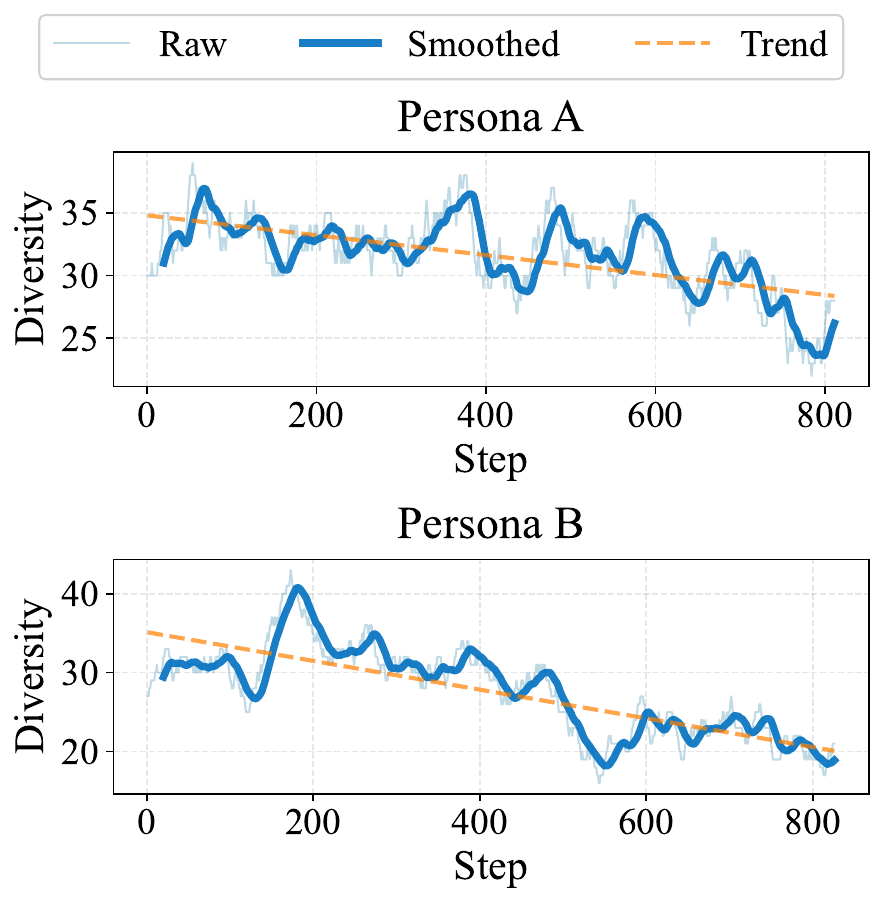}
\caption{Content diversity over interaction.}
\label{fig:diversity}
\end{figure}

\textbf{Bubble Breadth (Diversity).} 
We deploy agents on Bilibili with fresh accounts, and classify recommended videos 
into hierarchical categories while interacting with the platform. 
We track the number of distinct categories within sliding windows 
of 50 videos. As Figure~\ref{fig:diversity} shows, content diversity decreases 
over steps, quantifying how the recommendation algorithm narrows content 
exposure. Category entropy exhibits similar declining trends. 

\textbf{Observations.} After 800 interactions, Persona A exhibits a 20\% 
decrease in content diversity, while Persona B shows 
a more pronounced 40\% decline. Both reductions are significant, 
but Persona B's steeper decline correlates with its more distinct 
viewing patterns. Training data reveals that Persona A's 
watch duration distribution is concentrated, whereas Persona B 
exhibits a dispersed, bimodal distribution, indicating more 
polarized engagement behaviors.

\textbf{Bubble Depth (Bubble Escape Potential).} 
We measure escape potential via a \emph{sequential within-user counterfactual} 
experiment. 
Specifically, we deploy the reversed persona after the platform has 
adapted to the original persona. Each persona interacts with the platform 
for 400 steps, and we compute the content distribution divergence 
via Jensen-Shannon divergence reported in Table~\ref{tab:bep}.

\begin{table}[h]
\centering
\caption{Bubble Escape Potential (BEP) across platforms. Higher BEP indicates stronger escape potential and lower algorithmic inertia.}
\label{tab:bep}
\begin{tabular}{lcc}
\toprule
\textbf{Persona} & \textbf{Platform} & \textbf{BEP$\uparrow$} \\
\midrule
\multirow{3}{*}{Persona A} & Bilibili & \textbf{0.3393} \\
 & Douyin (TikTok) & 0.2896 \\
 & Kuaishou & 0.2723 \\
\midrule
\multirow{3}{*}{Persona B} & Bilibili & \textbf{0.2721} \\
 & Douyin (TikTok) & 0.1149 \\
 & Kuaishou & 0.2375 \\
\bottomrule
\end{tabular}%
\end{table}

\textbf{Observations.} 
Bilibili exhibits the highest BEP among all platforms, 
and Persona A consistently achieves higher BEP than Persona B. 
Most notably, Douyin yields an exceptionally low BEP of 0.1149 for Persona B, 
suggesting that Persona B's bimodal distribution induces stronger algorithmic 
inertia. Filter bubble escape thus depends critically on the interplay 
between platform design and behavioral profiles.

\textbf{Case Study.}
Table~\ref{tab:case_persona_a} illustrates Persona A's category shifts 
on Bilibili across cultivation and counterfactual phases.
Upon behavioral reversal, Entertainment-Comedy drops from 
17.50\% to 6.25\%, while Music categories surge.
Although this redistribution occurs, the moderate BEP (0.3393) reflects 
algorithmic inertia, and several cultivation categories still persist, 
which indicates incomplete escape from established filter bubbles.

\begin{table}[htbp]
\centering
\caption{Top-10 categories for Persona A on Bilibili.}
\label{tab:case_persona_a}
\resizebox{\linewidth}{!}{%
\begin{tabular}{clclc}
\toprule
& \multicolumn{2}{c}{\textbf{Cultivation Phase}} & \multicolumn{2}{c}{\textbf{Counterfactual Phase}} \\
\cmidrule(lr){2-3} \cmidrule(lr){4-5}
\textbf{Rank} & \textbf{Category} & \textbf{\%} & \textbf{Category} & \textbf{\%} \\
\midrule
1 & Entertainment/Comedy & 17.50 & Music/Performance {\scriptsize\textcolor{red}{(\textit{NEW})}} & 9.75 \\
2 & Knowledge/Science & 7.00 & Music/Live Performance {\scriptsize\textcolor{red}{(\textit{NEW})}} & 8.75 \\
3 & Gaming/Action Adventure & 5.00 & Animation/Short Films {\scriptsize\textcolor{blue}{($\uparrow$5)}} & 6.75 \\
4 & Life/Daily Life & 4.75 & Entertainment/Comedy {\scriptsize\textcolor{gray}{($\downarrow$3)}} & 6.25 \\
5 & Entertainment/Parody & 2.75 & Gaming/Action Adventure {\scriptsize\textcolor{gray}{($\downarrow$2)}} & 5.25 \\
6 & News/Current Affairs & 2.75 & Animation/Anime News {\scriptsize\textcolor{red}{(\textit{NEW})}} & 3.75 \\
7 & Gaming/Shooting Games & 2.75 & Animation/ACG Merch {\scriptsize\textcolor{red}{(\textit{NEW})}} & 3.75 \\
8 & Animation/Short Films & 2.75 & Travel/Scenery {\scriptsize\textcolor{red}{(\textit{NEW})}} & 2.50 \\
9 & News/Social News & 2.75 & Music/MV {\scriptsize\textcolor{red}{(\textit{NEW})}} & 2.25 \\
10 & Gaming/MOBA & 2.50 & Animation/Anime {\scriptsize\textcolor{red}{(\textit{NEW})}} & 2.00 \\
\bottomrule
\end{tabular}%
}
\end{table}

%% file: common/6future.tex
\section{Future Work}
While PersonaAct enables black-box evaluation of recommender systems, several directions remain for future work:

\begin{itemize}
    \item \textbf{Scale and Diversity}: Expand dataset scale with more users and platforms such as TikTok, YouTube, and Instagram to improve generalization across scenarios.
    \item \textbf{Long-term Simulation}: Extend observation windows to study how filter bubbles evolve and solidify over time.
    \item \textbf{Intervention Studies}: Test mitigation strategies such as diversity injection and friction mechanisms, and quantify their effects on filter bubbles.
    \item \textbf{Comprehensive Metrics}: Develop evaluation frameworks covering content diversity, fairness, and user experience.
\end{itemize}

%% file: common/7con.tex
\section{Conclusion}

We propose PersonaAct, a framework that combines breadth and depth dimensions for auditing filter bubbles through persona-conditioned multimodal agents, achieving cost-efficient, privacy-preserving, and authentic analysis at scale. 
By combining automated persona synthesis via structured interviews 
with multimodal behavioral modeling and reinforcement learning, 
PersonaAct achieves substantial simulation fidelity improvements over generic LLM baselines. 
Counterfactual auditing across three major platforms reveals significant filter bubble effects 
through both content diversity decline and platform-dependent escape potential. 
We release the dataset, collection tools, and implementation 
to support reproducible auditing of recommendation algorithms.

\section{Ethics Statement}

We collected data with informed consent and fully anonymized all recordings. 
PersonaAct benefits society by enabling transparent auditing of recommendation algorithms 
to identify and mitigate filter bubbles. 
While simulation techniques carry misuse risks, 
we address this through anonymized data release and consent requirements. 
The benefits of transparent algorithmic auditing outweigh these potential risks.